\documentclass[twocolumn, 10pt]{article}
\usepackage{amsmath,amsfonts,epsfig} 
\usepackage{amssymb}
\usepackage{fancyhdr}
\newtheorem{theorem}{Theorem}

\numberwithin{equation}{section}
\newcommand{\TR}{\lambda}

\newcommand{\TIM}{\ensuremath{T_{\rm max}}}

\newcommand{\Var}{{\rm Var}}

\newcommand{\field}[1]{\mathbb{#1}}

\let\Pr\Prob

\newcommand{\Temp}{T_{\rm abs}}

\title{SYNCHRONOUS RELAXATION FOR 
\\
PARALLEL ISING SPIN SIMULATIONS}
\author{\ Boris Lubachevsky \ \ \ \ \ \ \ \ \ \ \ \ \ \ \ \ \ \ \ \ \ \ \ \ \ \ \ \ \ \ \ \ \ \ \ \ \ \ \ \ \ \ \ \ \   Alan Weiss \\
{\em \ bdl@bell-labs.com \ \ \ \ \ \ \ \ \ \ \ \ \ \ \ \ \ \ \ \ \ \ \ \ \ \ \ \ \ \ \ \ \ apdoo@bell-labs.com}\\
\ Bell Labs, 600 Mountain Avenue, Murray Hill, New Jersey, USA}
\begin{document}
\date{}
\maketitle
\begin{abstract}
A new parallel algorithm for simulating 
Ising spin systems is presented.
The sequential prototype is
the $n$-fold way algorithm \cite{BKL75},
which is efficient but is hard
to parallelize using conservative methods.
Our parallel algorithm is optimistic.
Unlike other optimistic algorithms, e.g., Time Warp,
our algorithm is synchronous. It also belongs
to the class of simulations known as ``relaxation'' \cite{CS89};
hence it is named ``synchronous relaxation.''
We derive performance guarantees for this algorithm.
If $N$ is the number of PEs, then
under weak assumptions we show that
the number of correct events processed 
per unit of time is, on average, at least of order $N/\log N$.
All communication delays, processing time, and busy waits
are taken into account.


\end{abstract}
\section{Introduction}
The Ising model of computational physics was
introduced in 1925 \cite{I25}
for describing magnetization phenomena.
It has since been in use both for its original purpose,
and as a computational metaphor
in areas ranging from economics \cite{LLRM95}
to wireless communications \cite{BGKKLS97, EGLW93}
to material science \cite{NPW93}.

In a simple version of the model we have
a planar lattice of atoms, 
each of which may be in one of two spin states, $+1$ or $-1$.
The atoms flip their spins stochastically in a way that depends only 
on the states of their nearest neighbors, on an external magnetic field, 
and on temperature.

Ising spin simulations
have always been slow.
There have been a number of proposals 
for speeding them up.
The fastest available serial algorithm that does not violate
the stochastic properties of the procedure by Metropolis et al.
\cite{MRRTT53} is 
the $n$-{\em fold way} algorithm \cite{BKL75}.
A parallel version 
of the $n$-fold way was proposed in
\cite{L87, L88} 
and implemented in \cite{K98}.
Its speedup 
with respect to the serial $n$-fold way algorithm was not very high
in low temperature.
This was because at low temperature the algorithm performed
many
tentative calculations of flips that were rejected at the end.
In the procedure 
atoms along the boundaries of regions hosted by different processing elements
were subject to such rejections; 
only atoms in the interior of regions were not.
As the rejection rate asymptotically 
approaches 100\%, 
even if the fraction of atoms on the 
boundaries is small,
the rejection overhead becomes the dominant bottleneck in simulation.
Note that the main advantage of the $n$-fold way
over the procedure in \cite{MRRTT53} is absence of rejections.

The parallel algorithms in \cite{L87, L88}
are {\it conservative}.
Herein we describe a new algorithm for Ising spin simulations
whose aim is to eliminate the drawbacks associated
with rejections.
The algorithm is {\it optimistic;\/} 
it belongs to the class of ``relaxation" simulations \cite{CS89}.
It is also a synchronous algorithm,
and as such, it differs from Time Warp \cite{J85}.
Our ``synchronous relaxation'' implements 
the $n$-fold way algorithm in parallel
by repeatedly generating a time segment of the spin flip trajectory.
After each iteration, the PEs 
exchange information they generated, and correct the errors thus revealed.
Unlike conservative schemes in \cite{L87, L88},
computations are not wasted on rejections.
And unlike Time Warp, 
erroneous events are corrected in
a synchronous, iterative fashion rather than asynchronously.
However, iterations might introduce extra computations.

Since Time Warp is asynchronous, and our algorithm is not,
the behavior and analyses of these algorithms are quite different.
In our analysis
we estimate the number
of iterations needed to correct mistakenly simulated events
and the amount of work in each iteration.
Under weak assumptions, 
we show that the former is small and the latter is not excessive.
Synchronous relaxation was discussed in \cite{EGLW93} in a different context, 
and without formal proofs.

\thispagestyle{plain}
Note that Metropolis algorithm in \cite{MRRTT53} simulates a certain Markov Chain.
Markov Chain uniformization is put forward in \cite{HN95, NH93}
as a tool for parallelizing a sequential Markov Chain simulation.
Algorithms in \cite{L87, L88} turn out to be 
a specialization of more general schemes in \cite{HN95}.
It should be noted that
the actual development of Ising spin
simulations did not follow this logic.
Specifically, the rejection scheme \cite{MRRTT53},
was proposed as an
inherently serial algorithm,
without any reference to parallel processing or uniformization.
Then scheme \cite{BKL75} was proposed as a derivative of, and
an improvement over \cite{MRRTT53},
since it got rid of event rejections.
Note that according to \cite{HN95, NH93} it is
the scheme in \cite{MRRTT53} that would rather be considered
a derivative of \cite{BKL75},
the derivative which is
more amenable to parallelization.
Actually,
it took a while 
\cite{L87, L88}
to break the tradition of thinking the scheme \cite{MRRTT53}
as being inherently serial.
Optimistic methods of parallelization of Markov Chain simulations
are discussed in \cite{NH93}, but still in connection
with uniformization, i.e., with event rejections.
Since our present synchronous relaxation method 
for parallelization of
the scheme in \cite{BKL75} eliminates 
rejections, it apparently does not follow
from \cite{NH93}; 
its efficiency analysis is certainly new.
\section{The Ising model}\label{s:Isingmodel}
In the model atoms are located at the vertices 
of a rectangular subset $G$ of an orthogonal lattice.
A configuration is defined 
by the spin variables $s(v) = \pm 1$ for atoms $v\in G$.
In accordance with \cite{MRRTT53},
the time evolution of the configuration 
is a sequence of single spin updates: 
given a configuration, define the next configuration by choosing a 
vertex $v$ uniformly at random and changing $s(v)$ to $-s(v)$ 
with independent probability $p$, computed as described in the next paragraph.
With probability $1-p$, the $s(v)$ remains unchanged.

Computing the probability $p = p(v)$ involves
the 4 nearest neighbors of vertex $v=(i,j)$: 
$(i+1,j),(i-1,j),(i,j+1),(i,j-1)$.
Periodic boundary conditions are assumed in both directions.
Computing the $p(v)$ also involves the Hamiltonian
\begin{equation}
\label{hamil}
E = -J \sum_{v,v'\in G} s(v) s(v') - H \sum_{v} s(v) 
\end{equation}
where $J > 0$ and $H$ are constants, and the pairs $(v,v')$ of indices in the first sum
are nearest neighbors.
The value of $p$ is given in \cite{MRRTT53} by
\begin{equation}
\label{pMin}
p = \min \{ 1, \theta \}, \ \
\theta = \exp (- \Delta E /k_B \Temp )
\end{equation}
where 
$\Delta E = \Delta E(v)$ is the energy change that would result from
the flip of the spin at site $v$, $k_B$ is Boltzmann's constant, 
and $\Temp$ is the absolute temperature.
In \cite{G63} a different expression for $p$ is used:
\begin{equation}
\label{pglauber}
p = \theta /(1+ \theta )
\end{equation}
with the same $\theta$.
Note that computing $\Delta E (v)$ for spin $v$ involves only 
the values of $s(v)$ and $s(v')$ for the nearest neighbors $v'$ of $v$.
According to \eqref{pMin}, 
a spin flip which results in lower energy,
$\Delta E < 0$,
certainly occurs, because in such a case,
$p = 1$;
but in \eqref{pglauber} $p < 1 $ always
and even the flips with $\Delta E < 0$ can sometimes be rejected.
\section{The $n$-fold way algorithm}\label{s:serialnfoldway}
We suppose that 
attempts of a flip by each atom constitute
an independent Poisson process\footnote{
In \cite{BKL75, MRRTT53} system state changes
are just a sequence.
For the sake of parallelization we 
need spin flips to be a Poisson process in continuous time.
This assumption is consistent with the models in \cite{BKL75, MRRTT53},
but is not mentioned there.}
with fixed rate, say $\TR$.
Then
the successful attempts 
is also a Poisson process,
but with a smaller rate
$\TR p$, $p = p(t)$, which, in general, varies in time. 

The $n$-fold way algorithm in
\cite{BKL75}
splits 
all atoms into $n$ classes,
where a class is defined as those atoms
with the same flip probability $p$,
given either by \eqref{pMin} or \eqref{pglauber}.
In the 2D Ising model 
there will be no more than $n=10$ classes: 
each atom may be either up or down, 
and its four neighbors have 5 possibilities 
(none up, one up,$\ldots$, 4 up), 
so there are $2\times 5$ possible configurations for each atom
For each class $k$ define
\begin{align}
\label{eq:nidef}
N_k &= {\text{number of atoms in class }} k\\
\label{eq:pidef}
p_k &= \Pr\left ( {\text{atom in class }} k {\text{ flips when chosen}} \right ).
\end{align}
\thispagestyle{plain}
Given that the Poisson rate of flips of a single atom of class $k$ 
is $\TR p_k$,
the combined rate of flips of all atoms is
\begin{equation}
\label{biglambda}
\Lambda = \TR \sum_k N_k p_k,
\end{equation}
and the combined process whose arrivals coincide with spin flips at any
atom also enjoys the Poisson property.
As spins flip, membership of atoms may change,
$N_k$ may vary in time $N_k = N_k (t)$,
and so may the Poisson rate of the combined process
$\Lambda = \Lambda (t)$.

It follows that given the time $\tau_{i-1}$ 
of the $(i-1)$th flip,
the time of the $i$th flip can be
generated as
\begin{equation}
\label{nextau}
\tau_i = \tau_{i-1} + \frac{-\log U_i}{\Lambda(\tau_{i-1})},
\end{equation}
where $U_i$ is the $i$th independent sample 
of a random (0,1) uniformly distributed variable.
The first spin flip time $\tau_1$ can be generated as in \eqref{nextau}
with $i = 1$ if we formally set $\tau_0 = 0$.

We should then select the atom whose spin is to flip.
This is done in three phases.
In Phase 1 we generate an independent random (0,1) uniformly
distributed variable $V_i$.
In Phase 2 we choose a class $k_*$ 
to which the atom to be flipped belongs.
This is done by linearly
scanning the sequence of classes $k = 1,2,...n$
while summing the weights of their
chances to be selected
\begin{equation}
\label{Wsum}
W(k) = \sum_{j = 1}^k \frac { N_{j} p_{j}} { \sum_i N_i p_i } .
\end{equation}
Once the inequality 
\begin{equation}
\label{Wstop}
W(k-1) < V_i \le W(k)
\end{equation}
is detected
for $k = k_*$, we stop.

In Phase 3 we choose
an atom in class $k_*$.
All atoms in each class are maintained in a linear order.
To determine the index of an atom in the order we take 
the normalized residual $R_i$ of the random
sample $V_i$. 
This is defined as
\begin{equation}
\label{nores}
R_i = \frac {V_i - W(k_* -1)}  {\frac { N_{k_*} p_{k_*}} {\sum_j N_j p_j} } , 
\end{equation}
where $k_*$ is the class index found in Phase 2.
The required index is then 
\begin{equation}
\label{vindex}
\mbox{access index} = \lfloor R_i N_{k_*}\rfloor + 1 .
\end{equation}

Note that each spin flip
changes the class membership of several atoms.
We omit here the discussion 
of an elaborate data structure \cite{BKL75}
which maintains the classes in the computer
memory so that the atoms in the classes are positioned
in the required linear order.
We note that it only takes $O(1)$ computations
to adjust this order at each spin flip. 
Experiments \cite{BKL75} confirm that 
at a low temperature,
the $n$-fold way algorithm
is much faster than the original Metropolis et al. algorithm
with rejections in \cite{MRRTT53}.
\section{Synchronous relaxation: a general formulation}
\label{S:algorithm}
Suppose that a simulation of a system
is to be performed
on a multiprocessor with $N$ processing elements (PEs).
Our procedure is to give
each PE a subsystem to host,
and have the PE produce the history of
that subsystem.

Each PE will keep track of
the simulated time before which all simulated
histories are known;
this quantity is called {\em committed time.}
Committed time increases in steps,
in synchrony among all PEs;
its value is common to all PEs.
Each step consists of several iterations.
At each iteration,
each PE produces a tentative or speculative history
of the subsystem it hosts
beyond the current committed time.
The PE extends its local history until
its local time reaches the committed time plus \TIM,
where \TIM\ is the step size of
committed time increases.
Since subsystems are, in general, connected,
in order to produce a correct local history, 
a PE needs to know the correct local histories of other PEs.
But they are not known, because
other PEs are in the same quandary.
If, by some miracle,
a PE knew the correct histories of all others,
then it would produce a correct history for itself;
when all PEs produce correct histories,
then the next committed time
will become committed time plus \TIM.
So all we have to explain is how to achieve this miracle.

\thispagestyle{plain}
The mechanism is by iterations.
During the first iteration,
each PE makes the simplest assumption about
the histories of the other PEs;
we call this the canonical assumption.
For example, it may assume that the states of the
other subsystems do not change.
This assumption will enable it to produce its own history.
After all the PEs generate local history 
for an additional \TIM\ units of simulated time,
they compare their histories.
This comparison is done in synchrony,
in the sense that the PEs
perform the comparison only after they have all
completed the previous task of generating their histories.
As a rule, there will be inconsistencies between
the assumed and actual (generated) histories of other PEs.
If so, the PEs need to perform more iterations.

During subsequent iterations,
if a PE needs to know the local history
of another PE, it uses that history
generated in the last iteration.
The goal of producing correct
histories at a step is achieved
if no PE detects any inconsistencies
between
the assumed and actual history of any other PE.
Once this happens,
all PEs increase committed time by \TIM,
and continue.

We now give a
representation of the synchronous relaxation algorithm.
In the outset of the simulation, committed time is set to zero.
The subsystems hosted by the PEs are set to their initial states.
Then each PE executes the following code.
The execution is asynchronous except for
the two 
``synchronize'' statements.
While executing a ``synchronize,'' each PE waits
for the other PEs to
reach the same statement.
\\
\\
\fbox{
\begin{minipage} {7.8cm}

DO (step)
\begin{enumerate}
\item Choose a step size $\TIM$ for committed\\
\hspace*{0.2in}
time increase.
\item Make canonical assumption about \\
other PEs' histories.\\
\\
DO (iteration)\\
\begin{enumerate}
\item Generate local history, starting with\\
\hspace*{0.2in}
committed time until\\
\hspace*{0.2in}
local time $\ge$ committed time $+ \TIM$.
\item Synchronize.
\item \label{l:checkstep}
Compare generated history of other PEs\\
\hspace*{0.2in}
with the corresponding assumptions.\\
\hspace*{0.2in}
If they differ, replace the current \\
\hspace*{0.2in}
assumptions
with the corresponding \\
\hspace*{0.2in}
histories.
\item Synchronize.\\
\end{enumerate}
UNTIL (iteration) all PEs detect \\
\hspace*{0.2in}
no difference in the comparison in step \ref{l:checkstep}.
\item Increment committed time by \TIM.\\
\end{enumerate}
UNTIL (step) committed time exceeds \\
\hspace*{0.2in}
a prespecified value.

\end{minipage}}

There are two issues that need to be addressed.
Will there ever come a time when no
inconsistencies are detected?
Another is the question of correctness.
Is the lack of inconsistencies equivalent
to correctness of the history?
Generally, the iterations will not terminate.
However, when applied to discrete event
simulations, under mild conditions we can show that
the iterations do terminate.
Also, under weak conditions,
we can show that lack of inconsistency
is equivalent to correctness (see \cite{LW01}).

The preceding algorithm turns out to be
applicable not only to deterministic simulations,
but to stochastic
simulations, too, using the following technique.
Sequences of (pseudo)random numbers are employed in
stochastic simulations.
We should treat these sequences
as deterministic.
To this end,
all we have to do is to {\em reuse}
the 
(pseudo)random numbers generated at an iteration,
during the course of the following iterations.
An extreme way of effecting this is to generate
all random numbers in a list before the rest of the simulation
begins.
This is a helpful way to think about
simulation, whether or not it turns out
to be a practical method in any instance.

The general description above may not
give enough details for a specific implementation.
Furthermore, it may not be clear how to
estimate the efficiency of the algorithm;
specifically, how to bound the number of iterations
in each step.
The next sections will fill in
these details for the Ising model.
\thispagestyle{plain}
\section{Synchronous relaxation for Ising simulation}
\label{s:synchIsing}
In this section we detail the synchronous relaxation algorithm
for Ising simulation using the $n$-fold way algorithm.
While simulations can be implemented correctly
even if different PEs run different serial algorithms
to generate local history, we are interested in high performance,
so we will concentrate on a parallelization 
of the most efficient serial algorithm.

We follow the algorithm outlined in Section \ref{S:algorithm}.
The set $G$ is partitioned into subsets $G_i$, $1\le i\le N$,
and each $G_i$ is hosted by a separate processing element,
PE${}_i$. 
Each $G_i$ corresponds to a subsystem in the general description
of Section \ref{S:algorithm}.
As before, the boundary of $G_i$ is
denoted $\partial G_i$.

Statement 1 of the general relaxation procedure 
requires a choice of step size \TIM\ for
each committed time increase.
In the simplest algorithm formulation,
we choose the {\em same} step value
to serve for all committed time increase steps,
\begin{equation}
\TIM = \frac{\log N}{\lambda \left |\partial G_i \right |},
\end{equation}
assuming that all the sizes of boundary regions
$\left |\partial G_i \right |$ are equal,
and where $\lambda$, as in Section \ref{s:serialnfoldway}
is the largest rate at which
any atom can flip.

Statement 2 of the 
general relaxation procedure
requires each PE to make 
a canonical assumption 
concerning its neighbors' histories.
We choose the assumption that
no spin flip occurs
in any neighboring atom during the initial committed time
increase step \TIM.

Statement (a) 
requires each PE at each iteration 
to generate a segment of local history.
The segment begins at the current committed time value,
let us call it $t_c$, and 
ends at time $t_c + \TIM$.
The history can be identified
with a sequence 
$t_c \le \tau_1 < \tau_2 < \ldots < \tau_m < t_c + \TIM$
of spin flip times and the corresponding sequence of atoms
$v_1 , v_2 , \ldots v_m $
that flip their spins at these times.
Because the procedure is complex and subtle,
our discussion will be lengthy.
We split it into several cases.

\thispagestyle{plain}
\noindent
{\bf{First iteration, first step.}}
Thanks to the canonical assumption,
the procedure executed by each PE 
in this case is almost a literal repetition 
of the sequential algorithm described
in Section
\ref{s:serialnfoldway}.
As in that algorithm, a sequence of pairs 
$(U_1,V_1), (U_2,V_2), \ldots, (U_m,V_m)$,
of independent samples of uniform (0,1) distributed random variables
$U_i$ and $V_i$, feeds the generation of the spin flip history.
The $U_i$ feeds formula \eqref{nextau}, which generates
$\tau_i$, given $\tau_{i-1}$ and given the previous system state.
The $V_i$ feeds the procedure that selects the class
of the atom $v_i$ to flip and selects the $v_i$'s index
according to formulas \eqref{nores} and \eqref{vindex}.
To compute $\tau_1$ 
using \eqref{nextau} for $i=1$ we formally assume $\tau_0 = 0$.
As in the sequential case, this
can be done, because the state
of the system at time $t_c = 0$ is known.
The computations continue for as long as $\tau_i$ computed
by \eqref{nextau} is smaller than $t_c + \TIM$.
The last $\tau_i$ that satisfies this condition becomes $\tau_m$.

Steps (b), (c), and (d), that follow are obvious.
Note that while comparing the generated histories with
the assumed histories,
as required in step (c), 
the PE pays attention
only at the single layer of atoms that border its region.
This is due to the specific neighborhood 
structure that consists of 4 neighbors 
which we have assumed in our Ising model
(see Section \ref {s:Isingmodel}).
If the geometry were different (e.g., more dimensions, leading
to more neighbors; or deeper penetration of influence)
then we might need to check more atoms.

After exchanging information about their produced histories in step (c), 
the PEs detect inconsistencies.
Any boundary flip generated at the initial iteration causes
inconsistencies in neighbors, because of the canonical
assumption that there were no flips in boundaries.
The communication in step (c) leads the PEs 
to have new assumptions about the histories of their neighbors,
in place of the canonical assumption.


\noindent
{\bf{Subsequent iterations, first step.}}
After the PEs synchronize in step (d),
they execute step (a) again.
The first important distinction of this parallel procedure from its
sequential prototype is that the feeding sequence
$ (U_1,V_1), (U_2,V_2), \ldots$ of random numbers 
at the subsequent iterations has to be
the {\em same} as at the first iteration.
No new random sample $U_i$ or $V_i$ in
place of the previously generated one should be produced.
This is in keeping with the notion given at the
end of Section \ref{S:algorithm} that these
numbers are viewed as
presimulated, and need to be used in order.

As in the first iteration, each {\em old} $U_i$ is employed
to generate $\tau_i$ and each {\em old} $V_i$ is employed
to select atom $v_i$ for the flip.
Because of different boundary conditions,
the resulting $\tau_i$ and selected $v_i$ will generally
differ from those computed at the previous iteration.

Moreover, the second main distinction between
parallel and serial algorithms
is that computing $\tau_i$, given $\tau_{i-1}$
and given the system state at time $\tau_{i-1}$, is not
as straightforward as simply applying \eqref{nextau}.
The complication arises because spins in the neighborhood may flip 
during time interval $(\tau_{i-1} ,\tau_i)$, 
say at time $t_b$, which we call 
a {\em break point} time,
$\tau_{i-1} < t_b <  \tau_i$. 

The class membership of some atoms hosted
by the PE may change at time $t_b$ as a result. 
This in turn may change some $N_i$ and hence
may change the $\Lambda$ computed by \eqref{biglambda}
and used in \eqref{nextau}.
The presence of a break point $t_b$ inside interval 
$(\tau_{i-1} ,\tau_i)$, 
and the change of $\Lambda$ at $t_b$,
makes formula \eqref{nextau} unusable as it is.

A more general method 
adjusts $\tau_i$ so as to satisfy the following equation
\begin{equation}
\label{eq:lambdataua}
\int_{\tau_{i-1}}^{\tau_i} \Lambda (t)\, dt = -\log U_i .
\end{equation}
Solving \eqref{eq:lambdataua} for $\tau_i$
is easily accomplished since $\Lambda (t)$ changes
in time in a piecewise constant fashion.
One way of solving \eqref{eq:lambdataua} is,
with self-evident notation, to order
the break points from $\tau_{b,0} = \tau_{i-1}$ to
the maximum
$\tau_{b,m} = t_c + \TIM$, and to compute
for $1\le i\le m$ the sums
\begin{align}
\label{eq:breaksum}
b_k &= \sum_{j = 1}^k \left ( \tau_{b,j} - \tau_{b,j-1}\right ) \\
&= \int_{\tau_{i-1}}^{\tau_{b, k}} \Lambda(t)\, dt
\end{align}
If $-\log U_i > b_m$ then $\tau_i > t_c + \TIM$,
and we are done.
If $-\log U_i \le b_m$, then since
the $b_k$ are monotone increasing,
we easily find $k$ such that 
\begin{equation}
\label{eq:bsubk}
b_{k-1} < -\log U_i \le b_{k}.
\end{equation}
Given $k$, we have (analogously to equation \eqref{nextau})
\begin{equation}
\label{eq:newtau}
\tau_i = \tau_{b, k}+\frac{-\log U_i - b_{k-1}}{\Gamma(\tau_{b, k-1})}.
\end{equation}
Another, more efficient, way to solve \eqref{eq:lambdataua} for $\tau_i$,
is to generate the $b_k$ one at a time until equation \eqref{eq:bsubk}
is satisfied.

\thispagestyle{plain}
Another distinction from sequential simulation
is a possible need for {\em additional} random numbers.
The need may arise as a result of change of spin flip
times $\tau_i$ at iterations.
A subsequent iteration may fit
more spin flips, or fewer spin flips, within the 
time interval $(t_c, t_c + \TIM)$.
Recall that a presimulated list will have the
random numbers drawn in order, and so
additional random numbers must be drawn
after the original ones are used.
In particular, the random number $U_{m+1}$ that caused
$\tau_{m+1}$ to exceed $\TIM$ would be the first one used
if more random numbers are needed.
If fewer flips are needed, we save unused random numbers
for possible use in future iterations or subsequent time steps.

Iterations of the synchronous relaxation steps a, b, c, and d
continue until all PEs realize that the assumptions they
made at the beginning of an iteration agree with the information
they receive at the iteration's end;
that is, the times $\tau_i$ and flipped atoms $v_i$
are the same as in the previous-iteration assumption
and this hold for all
$\tau_i < \TIM$.
This signals that it is time
to advance the committed time to $t_c = \TIM$
for a new time step,
and to continue with a new round of iterations if necessary.

\noindent
{\bf{Subsequent steps.}}
The only difference of this situation from 
those already described is in defining initial conditions.
Whereas $\tau_0 = 0$ at the first time advancement step
and the initial system state 
is that at time 0,
for the following steps, the initial state
of the subsystem hosted by a PE
is the state resulting from the last simulated event
at the previous time advancement step.

Let $m$ be the index  of the last spin flip
simulated in the previous step.
Using the notation above, the last event if any of the
previous step was $(\tau_m ,v_m)$.
If, by chance, there were no events until now,
then $m = 0$, $\tau_m = 0$, and $v_m$ is undefined.
Redefining notations for the current time step,
the old $\tau_m$ becomes the $\tau_0$ for this time step.
This $\tau_0$ is used in this time step in the same
way as the $\tau_0$ of the first time step was used.
In particular, the $\tau_1$ can be found
using either formula \eqref{nextau} or
by solving equation \eqref{eq:lambdataua} with respect to $\tau_i$,
for $i = 1$.

What we called $U_i$ and $V_i$ at the previous time step
we now rename to $U_{i-m}$ and $V_{i-m}$;
so $U_{m+1}$ becomes $U_1$, etc.
Again, from viewing random
numbers as coming from a presimulated list,
any remaining random samples $V_{m+1}, U_{m+2}, V_{m+2},....$,
must be reused before additional random sampling is done.
They are to be renamed in the new time step as
$V_1, U_2, V_2,....$ respectively.

\noindent
{\bf{An alternate approach to subsequent steps.}}
For Ising simulations, properties of Poisson random variables
enable us to wipe clean the slate, empty event lists of all
PEs, and start afresh at the time $t_c + \TIM$.
That is, we need not keep any knowledge of tentative events
generated at previous steps.
New random numbers may be generated at this time, taken
later from a presimulated list than any up until now.
This procedure may not be desirable; we simply wanted to point out
that for Ising simulations it leads to unbiased simulations.
\section{Efficiency}\label{s:maineffic}
One of the nicest attributes of our algorithm is that it is provably efficient.
In this section we outline the assumptions that go into the proof,
and give a statement of the result.
The detailed proofs can be found in the forthcoming paper 
\cite{LW01}.

The efficiency result is asymptotic as $N$, the number
of PEs, becomes large.
Our method is to bound the probabilistic distribution
of the number of iterations needed to simulate one step
of the algorithm.
In addition, we bound the distribution of the amount of work
needed to simulate each iteration.
We also bound from below the average number of events simulated
in each step.
These bounds require a particular choice of $\TIM$ as
a function of $N$, and also require some
regularity of the model being simulated
as $N$ increases.
They also require that the load of simulated events 
remains balanced among the PEs as $N$ grows.

Specifically, suppose that the temperature is bounded away from
zero during the entire time interval being simulated,
and that $\lambda$ is bounded and is bounded away from zero,
so that the ratio between the largest and smallest Poisson rates
in the system remains bounded.
The blocks of PEs may change size as $N$ increases;
we suppose that the ratio between the number
of atoms hosted by the different PEs
remains bounded, that the ratio
of the number of boundary atoms between any
two PEs also remains bounded,
and that the number of neighbors
of any PE remains bounded,
all independent of $N$.
We also suppose that the number of boundary atoms per PE grows no
faster than $\log N$.
We choose $\TIM$ so that for each PE, the number of boundary
atoms times $\TIM$ is within a constant factor of $\log N$.
Clearly, this is possible because of the preceding assumptions.
This choice ensures that the maximum and minimum expected
number of events occurring among the boundary atoms of any
PE during a time interval of length $\TIM$ is of order $\log N$.

The preceding assumptions relate to the system being simulated.
Our efficiency result also requires some assumptions about
the machine doing the simulation.
We assume that synchronization can be done in no more
than order $\log N$ time.
We assume that communicating $x$ events from one PE
to a neighboring PE takes no more than a constant times $x + \log N$,
and that this can be done in parallel, so that if $x$ is the
maximum number of events to be communicated from any PE
to its neighbor, then all communication can be done in
no more than a constant times $x + \log N$ time.
We suppose that the speed of each PE is within a constant
factor of the speed of any other PE;
speed is the number of operations per
unit of time.
We suppose that the unit of time is chosen
so that the speeds of the PEs
is constant as $N$ increases.

Now for a bit of notation.
For any iteration, we let $f$ represent
the maximum total number of events that occur at the
boundary atoms of a PE.
That is, $f$ is the maximum number of events
that need to be communicated at each iteration.
We let $g$ denote the number of iterations
to complete one step.
Our assumptions show that the amount of time required to complete
a simulation cycle is bounded above by a constant times $fg$.
Note that  $f$, $g$, and hence $fg$ are random variables.
We let $K$ denote the maximum size of any set
consisting of a PE, its neighbors, and their neighbors.
$K$ is bounded by the assumption that the set of neighbors
is bounded in size.

We write $x\le_s y$ for real-valued random variables $x$ and $y$ if
\begin{equation}
\Pr(x>z)\le\Pr(y>z)
\end{equation}
holds for any real number $z$.

\thispagestyle{plain}
Under the assumptions delineated in the preceding section, we derive the following results.
\begin{theorem}
\label{thm:Ybound}
If the average number of events in a PE during an iteration is $\log N$,
then 
for each constant $C>1$,
\begin{equation}
f\le_s Ce\log N + Y,
\end{equation}
where $Y$ is a geometrically distributed random variable with parameter $1 / C$.
\end{theorem}
In particular, this theorem shows that $E(f) \le Ce\log N + 1/(C-1)$,
and $\sqrt{\Var (f)}\le Ce \log N + 2/(C-1)^2$.

We also find the following.
\begin{theorem}
\label{thm:geogbound}
If the average number of events in a PE during an iteration is $\log N$,
then for each constant $C > 1$ we find
\begin{equation}
g\le_s CeK\log N + X,
\end{equation}
where $X$ is a geometrically distributed random variable with parameter $1/C$.
\end{theorem}
In particular, this theorem shows that $E(g) \le CeK\log N + 1/(C-1)$ and
$\sqrt{\Var (g)}\le CeK\log N + 2/(C-1)^2$.

Combining these two theorems, we can bound the average time for a complete simulation cycle of length $\TIM$.
We may find a constant $c$ such that, for large enough $N$, by Schwarz's inequality,
\begin{equation}
\label{eq:fgebound}
E(fg)\le \sqrt{E(f^2)E(g^2)}\le (c\log N)^2.
\end{equation}
Since our assumptions show that the amount of time to complete
a simulation cycle is no more than proportional to $fg$, 
equation \eqref{eq:fgebound} gives our bound on the
mean real time required to simulate time $\TIM=\log N$.
Furthermore our assumptions show that in time $\TIM=\log N$
we expect to have to within a constant factor 
$N\TIM|G_i|\lambda p_i$ events simulated.
Therefore, the number of events simulated per unit time is, to within a factor $c$,
\begin{equation}
\frac{\text{\# events}}{\text{unit time}} = c\frac{N\log N}{(\log N)^2} = c\frac{N}{\log N},
\end{equation}
which is the efficiency estimate we promised.

We summarize our efficiency result as a theorem.
\begin{theorem}
\label{thm:MainResult}
Under the assumptions, the efficiency of the simulation 
is at least of order $1/\log N$ as $N\to\infty$.
Specifically, let $\tau(N, t)$ be the amount of real time
it takes to simulate an interval of time $(0, t)$ with $N$ processors.
Then there is a constant $B$ such that 
\begin{equation}
\frac{\tau(N, t)}{(t+1)\log N} \le B.
\end{equation}
\end{theorem}

We note that it can be shown that our scaling is optimal, in the sense that by
taking a function $\TIM = r(N)$ in place of $\log N$, where either
$r(N)/\log N \to \infty$ or $r(N)/\log N \to 0$ as $N\to\infty$,
gives strictly worse efficiency.
The details of all these results are in the full paper \cite{LW01}.
\thispagestyle{plain}

\end{document}